\documentclass[pdflatex,sn-mathphys-num]{sn-jnl}
\usepackage{graphicx}%
\usepackage{multirow}%
\usepackage{amsmath,amssymb,amsfonts}%
\usepackage{amsthm}%
\usepackage{mathrsfs}%
\usepackage[title]{appendix}%
\usepackage{xcolor}%
\usepackage{textcomp}%
\usepackage{manyfoot}%
\usepackage{booktabs}%
\usepackage{algorithm}%
\usepackage{algorithmicx}%
\usepackage{algpseudocode}%
\usepackage{listings}%
\usepackage[T1]{fontenc}
\usepackage[utf8]{inputenc}
\usepackage{siunitx}

\usepackage{bm}

\usepackage{subfig}

\usepackage{hyperref}
\hypersetup{
  colorlinks=true,
  linkcolor=blue,
  citecolor=blue,
  urlcolor=blue
}

\DeclareRobustCommand{\revchange}[1]{\textcolor{black}{#1}}

\DeclareSIUnit\angstrom{\text {Å}}

\newcommand{\conf}[3]{#1${#2}^{{#3}}$}

\begin{document}

\title[Article Title]{X-ray and extreme-ultraviolet spectra from collisions of Ar$^{18+}$ and O$^{8+}$ ions with neutrals}

\author*[1]{\fnm{Stepan} \sur{Dobrodey}}
\author[1,2,3]{\fnm{Chintan} \sur{Shah}} 
\author[1,4,5]{\fnm{Sonja} \sur{Bernitt}} 
\author[6]{\fnm{Ming Feng} \sur{Gu}} 
\author[7]{\fnm{Liyi} \sur{Gu}} 
\author[1]{\fnm{Thomas} \sur{Pfeifer}}
\author[1]{\fnm{José R.} \sur{{Crespo López-Urrutia}}}\email{crespojr@mpi-hd.mpg.de}

\affil*[1]{\orgname{Max-Planck-Institut für Kernphysik}, \orgaddress{\street{Saupfercheckweg 1}, \city{Heidelberg}, \postcode{69117}, \country{Germany}}}

\affil[2]{\orgname{NASA Goddard Space Flight Center}, \orgaddress{\street{8800 Greenbelt Rd}, \city{Greenbelt}, \postcode{20771}, \state{MD}, \country{USA}}}

\affil[3]{\orgname{Department of Physics and Astronomy, Johns Hopkins University}, \orgaddress{\street{3400 N. Charles Street}, \city{Baltimore}, \postcode{21218}, \state{MD}, \country{USA}}}

\affil[4]{\orgname{Helmholtz-Institut Jena}, \orgaddress{\street{Fröbelstieg 3}, \city{Jena}, \postcode{07743}, \country{Germany}}}

\affil [5]{\orgname{GSI Helmholtzzentrum f\"ur Schwerionenforschung}, \orgaddress{\street{Planckstra{\ss}e 1}, \postcode{64291}, \city{Darmstadt}, \country{Germany}}}

\affil[6]{\orgname{Space Science Laboratory, University of California}, \orgaddress{\street{7 Gauss Way}, \city{Berkeley}, \postcode{CA 94720},  \country{USA}}}

\affil[7]{\orgname{SRON, Space Research Organisation Netherlands}, \orgaddress{\street{Niels Bohrweg 4}, \city{Leiden}, \postcode{2333},  \country{The Netherlands}}}

\abstract{We present measurements of K-shell x-ray emission following  charge exchange of fully ionized argon with various neutral gaseous targets at small collision energies inside an electron beam ion trap. We also resolve the principal quantum number of electron capture in extreme-ultraviolet spectra from initially bare and hydrogen-like oxygen ions held in the same trap. We analyze discrepancies between these as well as previous measurements with theoretical models based on the multichannel Landau-Zener approach.
}


\maketitle

\section{Introduction} \label{sec:intro}
Charge exchange (CX) is a semi-resonant atomic process in which one, or several, bound electrons of a donor atom or molecule are captured by an ion during a collision. For highly charged ions (HCI), the  cross sections for this process far surpass those of radiative capture of free plasma electrons. CX collisions tend to populate highly excited states having binding energies close to the one of the donor species. Those excited states subsequently decay, either by autoionizing Auger processes, by emission of one or several photons, or combinations of these pathways. In many cases, the dominant channel is single-electron capture, and quite often a single X-ray photon is observed. However, radiative cascades to the ground state are also common. Photon emission following CX is of diagnostic interest for the astrophysical community; well-known examples are the X-rays from interactions of solar-wind ions with neutrals species in the solar system, in particular around comets~\citep{1996Lisse, 1998Owens, 2001Lisse, 2004Krasnopolsky}, planetary exospheres~\citep{2012Koutroumpa, 2006Dennerl, 2007Raymont, 2000Waite}, and the Moon~\citep{2004Wargelin}. Furthermore, CX is also important in MK-hot astrophysical environments, particularly in galactic winds~\citep{2012Liu, Fukushima2024}, and galaxy clusters~\citep{2015Gu, 2018Gu, 2018bGu, 2016Shah}. X-rays emitted due to CX can also provide insight into the shock physics of supernovae remnants and the element-abundance problem~\citep{2011Katsuda, 2012Katsuda, 2014Cumbee, 2019Uchida}. 

This process also drew strong attention due to the hypothesized presence of an unidentified weak spectral feature at $3.55\,\mathrm{keV}$ energy in the stacked X-ray spectra from various galaxies and galaxy clusters~\citep{2014Bulbul,2014Boyarsky}. A small excess at that energy was attributed to the decay of a sterile neutrino~\citep{2001Abazajian}, a hypothetical dark matter particle, since it could not initially be assigned to any tabulated atomic lines. Soon after, \citet{2015Gu} explained it based on expected CX processes in hot intracluster media inducing a transitions at that energy. Subsequently, a dedicated experiment by~\citet{2016Shah} showed how the predicted cluster of CX-excited transitions of hydrogen-like sulfur at $3.47(6)\,\mathrm{keV}$ could match the claimed X-ray excess. It is important to note that the modeling of the Ly-series transitions $np\rightarrow 1s$ for $n\leq4$ in S$^{15+}$  in the works of \citet{2014Bulbul,2014Boyarsky} can shift the centroid of the excess feature enough to explain the small remaining discrepancy.   
% %Incorporating this mechanism in the astrophysical plasma model reduces the residual of this line below $1\sigma$~\citep{2018Gu}. 
%
Similarly, it was suggested~\citep{2010Prokhorov} that an excess at 8.7~keV in the spectrum of the Galactic center observed by \textit{Suzaku} could be caused by the decay of a 17.4-keV mass sterile neutrino. Again, this could also reasonably follow from CX by Fe$^{25+}$ ions exciting the $\mathrm{Ly}\gamma$ line~\citep{2005Wargelin}. 

These two examples show how  understanding CX in thermal plasmas is a prerequisite for reliable interpretation of weak signatures in astrophysical X-ray spectra \citep{GuShah2023}.
Unfortunately, CX theory remains challenging, and several approximations are needed for modeling it. Including multi-electron capture (MEC) is particularly difficult \citep{Liang2021}. Moreover, the scarcity of experimental data, and in particular, of state-selective observation makes benchmarking existing models difficult. 

Although various experiments using electron beam ion traps (EBITs) showed good agreement with predicted cross-sections for electron capture into specific principal quantum number states $n$, cross-section calculations for angular momentum states $\ell$ for a given $n$ are inconclusive~\citep{2000Beiersdorfer, 2005Wargelin}. % , which is assumed to depend on the collision energy, is challenging~\citep{2014Martinez}.
Spectra of K and L shell transitions obtained with a high-resolution X-ray microcalorimeter showed even more striking inconsistencies with the theory~\citep{2010Leutenegger, 2013Leutenegger,2018Martinez}, as well as with widely used astrophysical models such as \textsc{SPEX-CX}~\citep{2016Gu} and \textsc{ACX}~\citep{2014Smith}. % models regarding the intensity ratios and energies of nickel L-shell transitions
These discrepancies could be caused by inaccurate transition energies, uncertainties in scaling equations used for estimating the $n\ell$-state in which the electron is captured, or inappropriate utilization of configuration mixing in the models, as explained in Refs.~\citep{2018Martinez,2017Cumbee}. 
It is important to note that CX also generates observable lines in the ultraviolet and optical range, as the CX-populated Rydberg levels stepwise decay through radiative cascades to the ground state. High-resolution spectroscopy at tokamaks showed the presence of strong extreme-ultraviolet (EUV) CX emission lines that are otherwise only weakly excited by direct electron impact~\cite{2015Beiersdorfer,2015Lepson}. 
Thus, an investigation of EUV photons emitted through radiative cascades starting from excited states after CX could help understanding how the $n\ell$-states are populated.

In the present work, we simultaneously measured in the X-ray and EUV wavelength ranges line emissions following CX  between fully ionized Ar and O ions and various neutral targets using an electron beam ion trap. We note some striking discrepancies to previous CX measurements, as well as departures from CX models based on the multichannel Landau-Zener (MCLZ) approach, such as Kronos~\citep{2016Mullen}, FAC~\citep{2008Gu}, and from X-ray spectral codes such as \textsc{SPEX-CX}~\citep{2016Gu}, and \textsc{ACX}~\citep{2014Smith}.

\section{Theory}
A straightforward description of CX is the classical over-the-barrier (COB) method~\citep{1980Ryufuku, 1986Niehaus}, where the combined electrostatic potential barrier of the highly charged projectile and the neutral target decreases during the mutual approach, leading to charge transfer at a certain internuclear separation. With this and other numerical methods like the classical trajectory Monte Carlo (CTMC) method~\citep{1966Abrines, 1977Olson}, total and $n$-resolved cross-sections can be predicted. More complete approaches like the multichannel Landau-Zener (MCLZ) approximation~\citep{1983Janev}, atomic-orbital close-coupling (AOCC), and quantum mechanical molecular orbital close coupling (QMMOCC)~\citep{1991Fritsch, 1993Shimakura} also yield the fractional populations of $nlS$ states for the captured electron. The MCLZ approach is widely used due to its comparatively low computational complexity and qualitative agreement with many experiments. Multi-electron capture, which becomes noticeable at lower collision energies typically prevalent in electron beam ion traps, is not included in most of the models.

Here, we evaluate MCLZ predictions of $n\ell S$-resolved cross-sections for single-electron capture.
The method calculates the transition probabilities between initial and final states in the collision. Already during the approach of the projectile with charge $q_1$, the neutral becomes polarized causing an attractive potential $V_i$ dependent on the internuclear distance $R$ between both partners given by
\begin{equation}
V_i = A\exp{(-BR)}-\frac{\alpha q_1^2}{2R^4},
\end{equation}
where $A$ and $B$ are coefficients estimated in~\citet{1980Butler} and the polarization is represented by the second term with the dipole polarizability $\alpha$ of the neutral target. After electron transfer, the interaction potential $V_f$ becomes repulsive
\begin{equation}
V_f = \frac{(q_1-1)q_2}{R} -\Delta E_{\infty}
\end{equation}
since both collision products have positive charge. In single-electron transfer, the charge $q_1$ of the projectile is reduced by one, and the neutral target becomes an ion with charge $q_2=1$. Then, $\Delta E_{\infty}$ represents the energy released by the charge exchange process, as the difference in the ionization potentials of projectile and target, including the excitation energy of the projectile after capture. The internuclear distance $R_C$ at which the charge transfer takes place is at the avoided curve crossing of the initial and the final potential curves. At the curve crossings $R_C$, the Olson-Salop-Taulbjerg~\citep{1976Olson, 1986Taulbjerg} equation yields:
\begin{equation}
\Delta V = \left(\frac{9.13f_{n\ell}}{\sqrt{q_1}}\right)\exp{\left(-\frac{1.324R_Ca}{\sqrt{q_1}}\right)},
\end{equation} 
where $f_{n\ell}$ is a correction factor for electron capture into non-degenerate $\ell$-states. The factor $a$ is given by $a=\sqrt{2I_P}$, where $I_P$ represents the ionization potential of the neutral~\citep{1986Taulbjerg}. The total transition probability is calculated by inclusion of all possible final states.
For bare-ion projectiles, $\ell$ states of same $n$ are degenerate, and thus, only the angular distribution of emitted photons gives information on the $\ell$ values of the cross-sections provided by MCLZ. The choice of the $\ell$ distribution function depends on the relative collision energy of the projectile and target. For comparatively high collision energies above $10\,\mathrm{keV}/\mathrm{amu}$, the statistical $\ell$-distribution described in \citet{1985Janev, 2004Krasnopolsky} with the relative $\ell$-populations
\begin{equation}
W_{n\ell}^{\mathrm{stat}} = (2\ell+1)/n^2
\end{equation}
is commonly used.
For lower values in the range of \revchange{$10$--$100\,\mathrm{eV}/\mathrm{amu}$}, both the low-energy weighting function
\begin{equation}
\label{eq:lowe}
W_{n\ell}^{\mathrm{low-e}} = (2\ell+1)\frac{[(n-1)!]^2}{(n+\ell)!(n-1-\ell)!}
\end{equation}
and the separable distribution
\begin{equation}
\label{eq:sep}
W_{n\ell}^{\mathrm{sep}} = \left(\frac{2\ell+1}{q_1}\right)\exp{\left(\frac{-\ell(\ell+1)}{q_1}\right)}
\end{equation}
are applicable.
Another well-established distribution is a modified low-energy function
\begin{equation}
\label{eq:mod_lowe}
W_{n\ell}^{\mathrm{low-e,Mod}} = \ell(\ell+1)(2\ell+1)\frac{(n-1)!(n-2)!}{(n+\ell)!(n-\ell-1)!},
\end{equation}
where intermediate angular momentum states of a given $n$ are preferentially populated, and the $s$-state ($\ell=0$) has a zero population. An explicit dependence on the collision energy is not given for these $\ell$-distributions.
Since these approaches are semi-empirical, they require experimental validation.

After calculation of $n\ell S$-resolved cross sections, a synthetic CX spectrum is generated using a radiative-cascade model. Electron capture populates highly excited states with relative fractions governed by these relative cross sections. Radiative stabilization into the ground state proceeds mostly through electric-dipole transitions with relative rates calculated according their branching ratios in all possible cascade channels.

\section{Experimental technique}
\begin{figure}
\centering
\includegraphics[width=0.8\columnwidth]{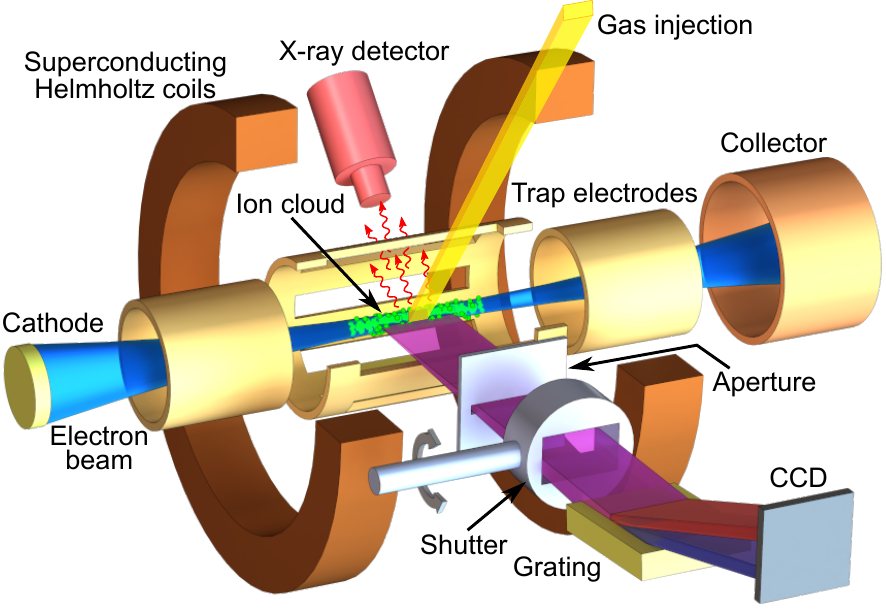}
\caption{\label{fig:EBITScheme}Scheme of the experimental setup. An electron beam emitted from a hot cathode is accelerated towards the trap center, where it crosses a tenuous atomic, or molecular, beam. HCI produced by electron impact become radially trapped by the negative space-charge potential of the electron beam, but their trajectories are also in part confined by the magnetic field. When the electron beam is turned off, the ions remain trapped by this field, as in a Penning trap. Photons emitted both in the electron-beam-on and magnetic-trapping modes are monitored in the X-ray and extreme-ultraviolet domains.}
\end{figure}

The measurements were performed in an EBIT applying the magnetic-trapping mode (MTM)~\citep{2000Beiersdorfer}, as in other works \citep{2005Wargelin, 2016Shah}. The device was equipped with a grazing-incidence spectrometer for the extreme-ultraviolet (EUV) and a commercial silicon-drift detector (SDD) for the X-ray range. Both instruments were mounted side-on with respect to the propagation axis of the electron beam.
A schematic of the setup is shown in Fig. \ref{fig:EBITScheme}. An electron beam with a current of $\sim200\,\mathrm{mA}$ originating from a thermoionic cathode is accelerated towards a stack of cylindrical trap electrodes, compressed to a diameter of approximately $50\,\mathrm{\mu m}$ in the trap center by a 6-T magnetic field generated by a pair of superconducting Helmholtz coils, and lastly dumped onto the inner wall of a water-cooled cylindrical collector electrode.

Neutral gas is injected using a differentially pumped setup giving raise to a pressure in the order of $10^{-10}\,\mathrm{mbar}$ at the trap center. This collimated atomic beam crosses the electron beam, leading to electron-impact ionization. The resulting positively charged ions are radially confined mainly by the negative space-charge potential of the electron beam, as well as axially by an electrostatic potential well formed by voltages applied to the outer trap electrodes. The resulting ion cloud typically containing $>10^7$ HCI has a length of $\sim50\,\mathrm{mm}$ and a diameter of $\sim500\,\mathrm{\mu m}$.

The cycle starts with the beam on. After ions in the desired charge state have been generated, the electron beam is switched off within microseconds using a fast high-voltage amplifier controlling the electron gun emission. This stops electron-impact excitation of the HCI, and subsequently photon emission. Moreover, radial trapping by the negative space charge of the electrons vanishes, and the ion cloud slowly expands to a diameter of $\sim1.5\,\mathrm{mm}$. The only processes causing emission of radiation are now the decay of metastable states and CX of HCI with neutrals. The former can be excluded by waiting some milliseconds before starting the spectral acquisition in the MTM period. Since we extend this over several seconds, metastable states decay becomes negligible. This on-off cycle is repeated continuously during the measurements.

EUV spectra are recorded  with a grazing-incidence grating spectrometer with a resolving power of $\lambda/\Delta\lambda \approx 400$ at $15\,\mathrm{nm}$ equipped with a windowless CCD camera. This instrument is mounted side-on and uses the trapped ensemble as its entrance slit, as shown in Fig. \ref{fig:EBITScheme}. To block photons emitted outside the MTM, a mechanical shutter is installed between the trap and the spectrometer. An arbitrary-function generator (AFG) triggers the switching of the electron beam and simultaneously controls the shutter. A synchronized linear ramp signal synchronized at the beginning of each cycle was fed into the data acquisition system to provide timing information by digitizing its value when a photon is detected. Hundred milliseconds after the electron beam is switched off and most metastables states have already decayed, the shutter opens, letting spectral integration on the CCD continue, and closes half a second before the electron beam was switched on again. Each CCD image integrates signal for $1800\,\mathrm{s}$, with approximately 200 switching cycles being accumulated. Background measurements ware obtained without injection of neutral gas, but with otherwise unchanged experimental conditions. After correcting for aberration and distortion of the spectral lines on the CCD, the obtained two-dimensional histograms are projected onto the dispersive axis of the spectrometer to generate the spectra. 

\begin{figure*}
\centering
\includegraphics[width=\textwidth]{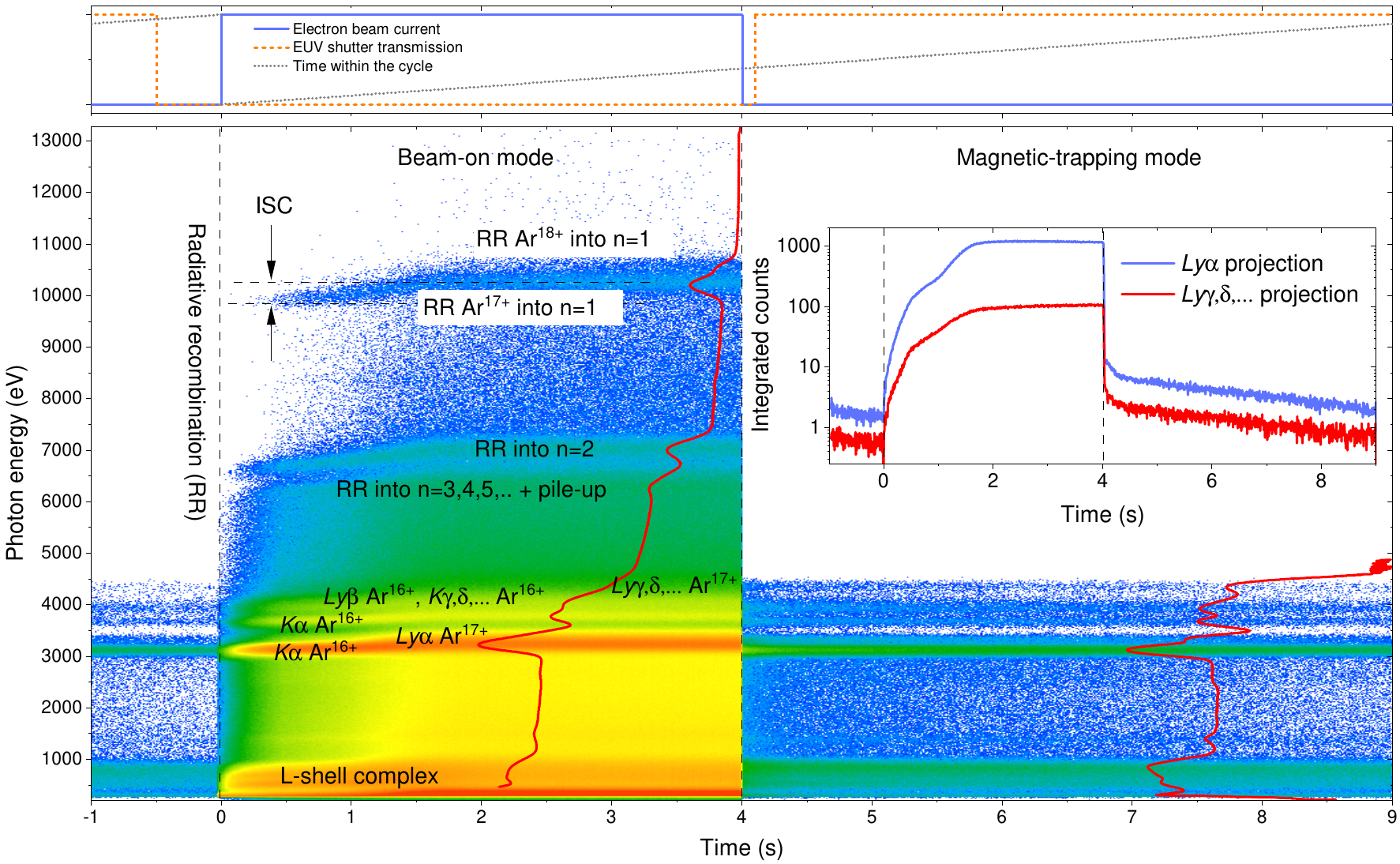}
\caption{\label{fig:ArCX2D} Two-dimensional histogram of photons emitted during electron-beam on mode and magnetic-trapping mode (MTM) of highly ionized argon up to the bare charge state interacting with neutral Ar atoms. The energies of the X-ray photons are plotted against the time within the switching cycle of the electron beam. Red curves display the total number of photon events projected onto the spectral axis during the beam-on and beam-off (MTM) times, respectively. An inset shows the same for the $Ly\alpha$ and $Ly\gamma,\delta,...$ transitions of $\text{Ar}^{17+}$ in the range from $3270\,\mathrm{eV}$ to $3470\,\mathrm{eV}$ and from $4270\,\mathrm{eV}$ to $4470\,\mathrm{eV}$, respectively, as a function of time. The top panel illustrates the switching sequence for the electron beam, the EUV shutter, and the linear time ramp. We label spectral features due to radiative recombination into open shells as RR; ISC marks the build-up of positive space-charge potential due to the filling of the trap with HCI during the breeding cycle.}
\end{figure*}

X-rays from radiative decay into the K-shell are recorded with a windowless solid-state silicon drift detector (SDD, model KETEK AXAS-M), mounted on the side of the trap. It has an effective area of $80\,\mathrm{mm}^2$ and sits at a distance of $141\,\mathrm{mm}$ from the trap center. The energy of each of the detected photons is recorded as a function of time within the switching cycle. A schematic of the switching procedure is shown in the top panel of Fig. \ref{fig:ArCX2D}.

Different gases are injected continuously to serve as educts for the highly charged ions and as neutral donors at the same time. For the study of CX between ions and neutrals from dissimilar elements, an additional, pulsed beam of molecular hydrogen is used. Also residual gas components present in the trap are studied as donors by turning off and on the external injection.

\begin{figure*}
\centering
\includegraphics[width=\textwidth]{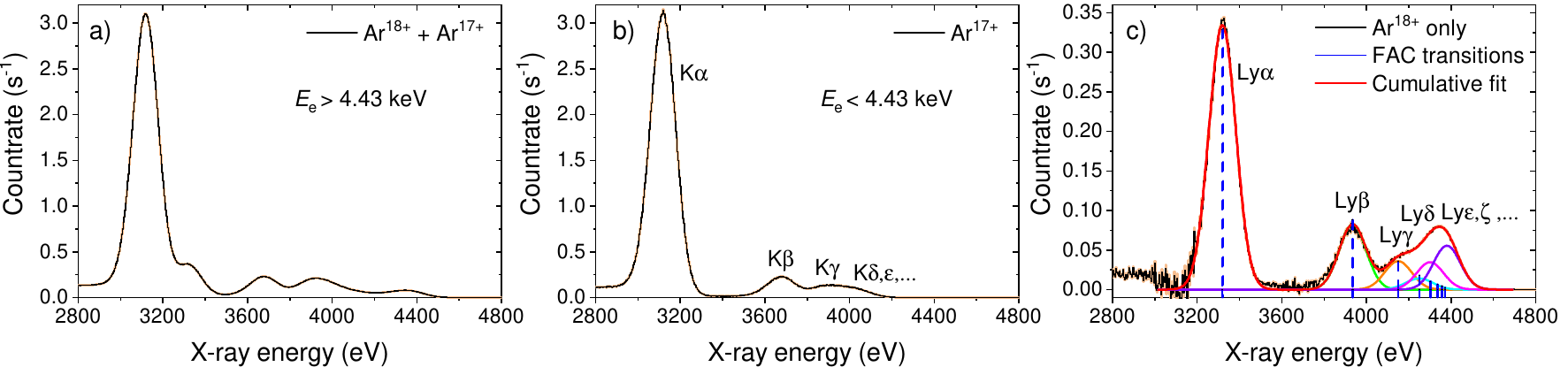}
\caption{\label{fig:NormalizationCX}CX-induced emission spectra in MTM of highly ionized argon colliding with neutral gas. \textbf{a)} $\mathrm{Ar}^{18+}$ and $\mathrm{Ar}^{17+}$ ions were produced at an electron-beam energy of $6\,\mathrm{keV}$, above the ionization threshold for the production of fully ionized argon. \textbf{b)} Only $\mathrm{Ar}^{17+}$ ions are produced at a beam energy below the ionization threshold at $4.42\,\mathrm{keV}$. \textbf{c)} Normalization of both spectra to the helium-like \revchange{$\mathrm{K}_{\beta}$ transition at $\sim3700\,\mathrm{eV}$}, which is present in both a) and b). Subtraction yields the pure contribution of CX of $\mathrm{Ar}^{18+}$ with neutral gas. The predicted hydrogen-like Lyman-transition energies are indicated by vertical blue dashed lines. A synthetic spectrum excited by mono-energetic electrons of $E_e = 6\,\mathrm{keV}$ is shown in green. Gaussians corresponding to the $n_{\mathrm{CX}}\to n=1$ transitions are fitted to the data.}
\end{figure*}

\begin{figure*}[h]
\centering
\includegraphics[width=0.65\textwidth]{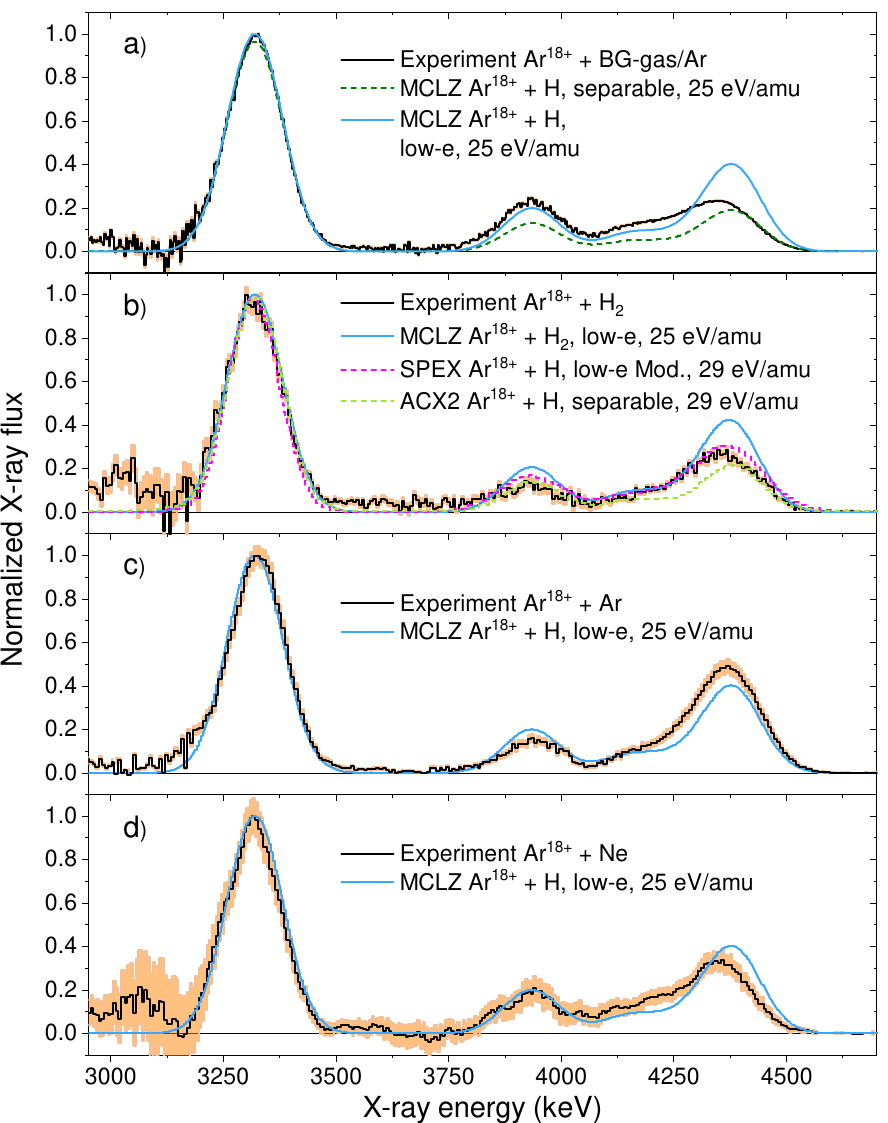}
\caption{\label{fig:Ar_CX_DifferentElementsGaussians}CX spectra of $\mathrm{Ar}^{18+}$ interacting with a mixture of residual gas and argon (a), with pulses of $\mathrm{H}_2$ (b), argon (c), and neon (d). The blue-colored curve represents synthetic charge-exchange spectra of $\mathrm{Ar}^{18+}$ with $\mathrm{H}_2$ in panel b), and with atomic hydrogen in the other panels. The calculations are based on the MCLZ approach with a low-energy $\ell$-distribution (Eq. \ref{eq:lowe}) at a relative collision energy of $25\,\mathrm{eV}/\mathrm{amu}$. The high compensation by ions of the space charge of the electron beam reduces the actual trapping potential substantially. After the electron beam is turned off, the ion-cloud expansion further reduces the mean kinetic energy of the HCI. The green dashed curve in a) shows a MCLZ calculation with the separable $\ell$-distribution of Eq. \ref{eq:sep}. All spectra are normalized to the Lyman-$\alpha$ transition. The orange-shaded region represents the statistical uncertainty.}
\end{figure*}

\subsection{Spectral calibration}
The X-ray detector was roughly calibrated with well-known K-shell transitions of hydrogen-like nitrogen, oxygen, neon and argon at $500.3\,\mathrm{eV}$, $653.6\,\mathrm{eV}$, $1021.7\,\mathrm{eV}$, and $3320.6\,\mathrm{eV}$, respectively, as calculated with  Flexible Atomic Code (FAC)~\citep{2008Gu}. These four gases were injected successively into the trap and ionized to bare and hydrogen-like charge states. Excitation to $n=2$ and subsequent radiative stabilization to the ground state by emission of Ly-$\alpha$ photons were observed with an energy resolution of $\mathrm{FWHM}\approx150\,\mathrm{eV}$. Gaussian distributions were fitted to the recorded spectrum for determining the line centroids. Transition energies calculated with FAC were assigned an estimated theoretical uncertainty of $10\,\mathrm{eV}$, but the present experiment does not require a better calibration. A linear regression determines the scaling factor between $4096$ detector channels and energy in electron volts.
%of $3.15\,\mathrm{eV/channel}$ with an offset of $19.5\,\mathrm{eV}$.

Calibration of the EUV spectrometer was performed by assigning the well-known wavelengths of hydrogen-like, helium-like, lithium-like and beryllium-like transitions of oxygen ions to the lines observed during the electron-beam-on mode. Gaussian peaks were fitted to the histograms obtained from the CCD images for determination of centroid positions in units of pixels. The wavelengths were taken from the NIST database~\citep{NIST}, and we used a second-order polynomial with corresponding residuals of less than $5\times10^{-3}\,\mathrm{nm}$ (compare Tab. \ref{tab:CXLines}) to fit the dispersion curve. \revchange{In magnetic-trapping mode, the ion cloud expands after the disappearance of the electron beam and its corresponding space-charge potential. Moreover, we observe that its mass center also shifts radially by about $\sim300\,\mathrm{\mu m}$. This displacement is attributed to the concomitant change in radial force balance, together with residual asymmetries of the electrostatic trapping potential and the magnetic field, and the magnetron-type motion of the ion cloud ensemble around the actual magnetic field axis. In our slitless geometry, where the narrow ion cloud is imaged on the detector, the shifted centroid changes the angle of incidence on the grating and produces a nearly wavelength-independent offset of $\sim0.02\,\mathrm{nm}$ over each $\sim5\,\mathrm{nm}$ spectral setting. All EUV spectra acquired in MTM are corrected for this displacement, and the resulting uncertainty included in the quoted value. Nonetheless, we cannot exclude that small offsets of individual CX features in Fig.~\ref{fig:O_CX_EUV} can be caused by unresolved satellite blends populated by multi-electron capture followed by autoionization, especially for collisions with molecular oxygen donors. We do not expect that our identification of the key EUV transitions is affected by this.}

\section{Results and data analysis}

\subsection{X-ray spectra}
A typical CX spectrum in the X-ray domain is shown in Fig.~\ref{fig:ArCX2D}. Here, argon atoms injected into the trap serve both as educts for the HCI production, and neutral donors for the CX process. The X-ray spectrum is shown as a function of the time within the switching cycle. At $t=0\,\mathrm{s}$, the electron beam is switched on, and fluorescence due to electron-impact excitation into various principal quantum number states with subsequent radiative stabilization into the ground state can be observed between $3\,\mathrm{keV}$ and $4.5\,\mathrm{keV}$. Furthermore, radiative recombination (RR) of mono-energetic free electrons from the electron beam into $n=1$ of $\mathrm{Ar}^{18+}$ and $\mathrm{Ar}^{17+}$ at $10\,\mathrm{keV}$ and into $n=2$ at $7\,\mathrm{keV}$ can be seen, where the energy of the emitted photons is the sum of the kinetic energy of the free electron and the binding energy of the state in which it is captured. The increasing slope of the RR is due to the filling of the trap with positive HCI, which partially compensates the negative space charge of the electrons. This results in a noticeable increase of the kinetic energy of the free electrons in comparison with the empty trap. This ionic space charge potential (ISC) changes the electron-beam energy at the trap by $\sim300\,\mathrm{eV}$ within two seconds. After switching off the beam at $t=4\,\mathrm{s}$, the interaction with free electrons stops, and the sole process resulting in X-ray emission is CX. Electron transfer into highly excited states of $\mathrm{Ar}^{17+}$ after capture into $n\geq3$ and subsequent relaxation into the ground state by emission of Lyman-$\gamma,\delta,\dots$ X-rays at $\sim4.4\,\mathrm{keV}$, Lyman-$\beta$ transitions from $n=3$ to $n=1$, and the final transition of the cascade, the Lyman-$\alpha$ transition at $\sim3.3\,\mathrm{keV}$ from $n=2$ to $n=1$ can be seen. The electron-beam energy of $6\,\mathrm{keV}$ is well above the threshold for the production of fully ionized argon. For atomic transitions into the ground state in the energy range between $3\,\mathrm{keV}$ and $4.5\,\mathrm{keV}$, a vacancy in the K-shell is required. Consequently, such X-rays can only originate from electron capture by bare and hydrogen-like ions, giving rise to K-shell transitions in hydrogen-like and helium-like ions, respectively. Since bare ions have two K-shell vacancies, CX can successively induce emissions of the two types.

To study CX by bare argon, the contribution of the hydrogen-like system has to be subtracted as described in \citet{2016Shah}. For this, a second measurement with an electron-beam energy of $4.4\,\mathrm{keV}$, thus slightly below the production threshold for $\mathrm{Ar}^{18+}$, was performed. A pure CX spectrum of helium-like argon then follows electron capture. After integrating all events in MTM onto the X-ray energy axis and normalization of the histograms to a well-separated $\mathrm{K}_{\beta}$ transition present in both spectra, the contribution of CX from $\mathrm{Ar}^{18+}$ is extracted by subtraction, as illustrated in Fig. \ref{fig:NormalizationCX}.

Charge exchange was also measured for collisions of $\mathrm{Ar}^{18+}$ with different neutral targets at similar trapping conditions, with an axial trap depth of $800\,\mathrm{V}$ and an electron-beam current of $200\,\mathrm{mA}$.
The results are presented in Fig. \ref{fig:Ar_CX_DifferentElementsGaussians} as solid black curves with corresponding statistical uncertainties in orange. 
%====================================================
In panel a), the donor is a mixture of argon and residual gas with a comparable fractional abundance. In panel b), the $\mathrm{Ar}^{18+}$ projectiles interact with a pulse of molecular hydrogen with a duration of $\sim100\,\mathrm{\mu s}$. Panel c) shows the pure CX spectrum of $\mathrm{Ar}^{18+} + \mathrm{Ar}$ with a negligible amount of residual gas in the trap. In d), argon is replaced by neon. Residual argon left from previous measurements in the vacuum chamber served to generate HCI for interacting with injected neon atoms. To estimate the fractional amount of residual argon and eliminate this contribution, the radiative recombination rates into $n=1$ of fully ionized argon and neon during the electron-beam-on mode, respectively, were compared. The corresponding fraction of CX events between $\mathrm{Ar}^{18+}$ and neutral argon is estimated to $14\,\%$ and subtracted by utilization of the pure $\mathrm{Ar}^{18+} + \mathrm{Ar}$ CX spectrum shown in panel c). In these measurements (b), c), and d)), the residual gas pressure measured between the 40-K thermal shield of the superconducting magnet and the vacuum chamber at room temperature was $<5\times10^{-10}\,\mathrm{mbar}$ corresponding to a neutral-target density of $< 1.6\times10^{5}\mathrm{cm}^{-3}$ at the $4\,\mathrm{K}$ cryogenic environment of the trap center. This contribution to the neutral-target composition is negligible compared to that of injected target gas, which is in the order of $10^{-7}\,\mathrm{mbar}$ in the first differentially pumped stages, and corresponds to a neutral density of $\sim 3\times10^{7}\,\mathrm{cm}^{-3}$ at the trap center. The ion density can be estimated from the slope of the radiative recombination of $\mathrm{Ar}^{17+}$ into $n=1$ in Fig. \ref{fig:ArCX2D}. At time $t=0$, the RR feature appears at a photon energy of $9825\,\mathrm{eV}$ with a nominal electron energy of $6020\,\mathrm{eV}$ and a beam current of $220\,\mathrm{mA}$. By subtracting the binding energy of $1s$ level into which the electron is captured, namely $4125\,\mathrm{eV}$, the actual free electron energy is found to be $5704\,\mathrm{eV}$. Thus, the contribution of the negative space charge of the electrons is $\approx 316\,\mathrm{eV}$. From the increase of the energy of RR photons caused by a nearly full compensation of the negative space charge by HCI, the total content of the trap can be estimated to $2\times10^8$ $\mathrm{Ar}^{16+}$ ions. With an estimated mean diameter of the ion cloud of $\sim380\,\mathrm{\mu m}$ and a length of $8\,\mathrm{cm}$ the ion density can be determined to $\sim5\times10^{9}\,\mathrm{ions}/\mathrm{cm}^{3}$.

\subsection{Comparison of X-ray spectra with predictions and previous experiments}
To simulate CX, we used several different spectral models, mostly based on the MCLZ CX cross sections. For the comparison presented in Fig.~\ref{fig:Ar_CX_DifferentElementsGaussians}, we generated synthetic spectra of $\mathrm{Ar}^{18+}$ interacting with $\mathrm{H}$ and $\mathrm{H}_2$ using the MCLZ-based Kronos code~\citet{2016Mullen}, where we used the low-energy distribution (Eq.~\ref{eq:lowe}) for a relative collision energy of $25\,\mathrm{eV}/\mathrm{amu}$ (solid blue curves). 
The green dashed curve represents the separable $\ell$-state distribution according to Eq.~\ref{eq:sep} calculated for the same parameters using Kronos.
Besides, we generated synthetic CX spectra using the models \textsc{SPEX-CX} (v3.05,~\citet{2016Gu}), and \textsc{AtomDB-ACX} (v2.0,~\citet{2014Smith}). Both incorporate the same $n$-resolved cross sections from the Kronos calculation, while the higher-order approximations (on quantum numbers $\ell$ and $s$) might differ. For the SPEX-CX model, we set the collision velocity to $50\,\mathrm{km}/\mathrm{s}$, and considered the recommended $n$-dependent weighting function on the $\ell$-distribution. In contrast, we used separable $\ell$-state distribution (Eq.~\ref{eq:sep}) in the ACX model. Here, the model uses the center of mass energy to obtain cross sections for the electron capture into each atomic shell from Kronos database. The curves in Fig.~\ref{fig:Ar_CX_DifferentElementsGaussians} show these synthetic spectra. Note that only single electron capture from atomic hydrogen target is considered in both models. 

For the comparison with other available CX measurements, we use the  hardness ratio $\mathcal{H}$~\citep{2001Beiersdorfer}, which provides indirect information on the relative collision energy. It is defined as the ratio of the intensity of hard X-rays to soft X-rays, in particular, the total flux $F$ of all $n\geq3\to n=1$ transitions and the Lyman-$\alpha$ transition:
\begin{equation}
\mathcal{H} = \frac{\sum_{i=3}^{\infty}F(n=i\to n=1)}{F(\mathrm{Ly}\alpha)}.
\end{equation}

We calculate $\mathcal{H}$ for all CX spectra by the ratio of the sum of the areas of the single fitted Gaussian distributions for all transitions $n\geq3\to 1$ and $n=2\to1$ taking the corresponding statistical uncertainties into account.

\begin{figure}
\centering
\includegraphics[width=0.85\columnwidth]{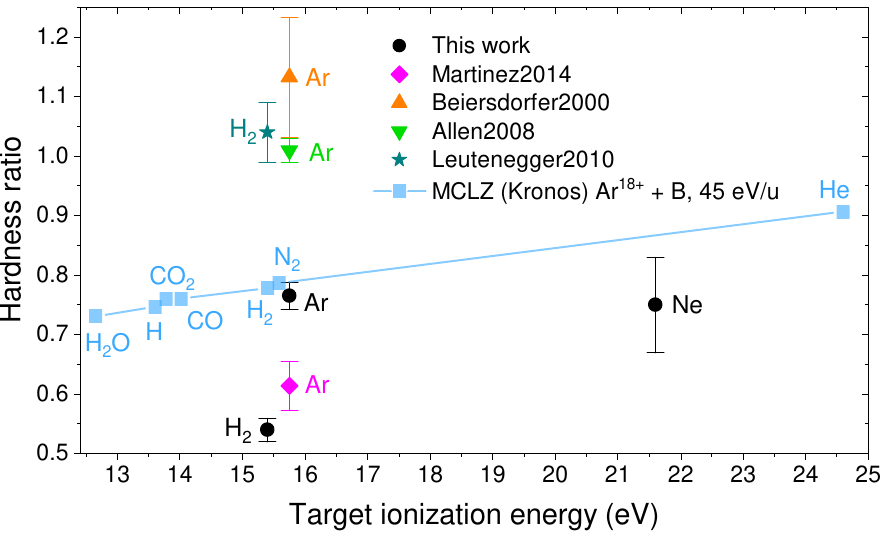}
\caption{\label{fig:HardnessIp}Hardness ratio of $\mathrm{Ar}^{18+} + \mathrm{A}$ as a function of the ionization potential of the neutral target species $\mathrm{A}$. The light-blue curve shows the prediction based on a MCLZ calculation with a low-energy $\ell$-distribution for a relative collision energy of $25\,\mathrm{eV}/\mathrm{amu}$.}
\end{figure}

In Fig. \ref{fig:HardnessIp}, the results for $\mathcal{H}$ for collisions of $\mathrm{Ar}^{18+}$ with different neutral targets from Fig. \ref{fig:Ar_CX_DifferentElementsGaussians} are shown as a function of the ionization potential of the target.
%====================================================================
No obvious scaling of $\mathcal{H}$ with the ionization potential $I_{\mathrm{P}}$ can be observed. The model based on an MCLZ approach with the low-energy distribution for the angular momentum states and a collision energy of $25\,\mathrm{eV}/\mathrm{amu}$ predicts an increase of $\mathcal{H}$ with higher $I_{\mathrm{P}}$. Furthermore, a significant discrepancy is observed between our measurements and others (\cite{2008Allen, 2014Martinez, 2000Beiersdorfer, 2010Leutenegger}) for $\mathrm{Ar}^{18+} + \mathrm{H}_2$, as well as for $\mathrm{Ar}^{18+} + \mathrm{Ar}$. 
The hardness ratio determined for the former collision in our experiment and in Ref.~\citep{2010Leutenegger} differs almost by a factor of two. 
\revchange{Moreover, for CX between $\mathrm{Ar}^{18+}$ and neutral argon, $\mathcal{H}$ determined by~\citet{2000Beiersdorfer, 2014Martinez} also departs by almost a factor of two, despite both measurements having been performed using the same EBIT at low collision energies $\leq25$~eV amu$^{-1}$.
It should be noted that the X-ray microcalorimeter employed in \citet{2014Martinez} had a considerably higher spectral resolution than the solid-state SiLi detector used in~\citet{2000Beiersdorfer}. The energy-dependent detector efficiency could in principle affect $\mathcal{H}$; however, both experiments accounted for the detector efficiency, including plausible ice contamination in the cryogenic X-ray microcalorimeter of~\citet{2014Martinez}. 
Finite detector resolution, quantum detection efficiency, line blending, and differing fitting procedures are therefore unlikely to account for this discrepancy on their own. A factor-of-two difference more plausibly reflects differences in the collision energy distributions, the presence of residual molecular or ionic donors, or possible multielectron capture processes. Further experiments are thus needed to understand such discrepancy.
}
%

%{because the ratio combines the Ly-$\alpha$ line near $3.3\,\mathrm{keV}$ with higher Lyman lines up to about $4.5\,\mathrm{keV}$. For the present windowless SDD this response changes smoothly over the relevant energy interval and cannot by itself account for a factor-of-two spread. However, different efficiency corrections, different spectral resolutions, and different treatment of unresolved high-$n$ blends may contribute to the inter-experiment scatter and should be considered when comparing absolute values of $\mathcal{H}$ from different EBIT measurements.}

To evaluate the effect of the trapping conditions, the interaction of $\mathrm{Ar}^{18+}$ with neutral argon is examined for different axial trapping potentials $V_0$, since this parameter determines the relative collision energy $E_{\mathrm{coll}}$. It can be described by the semi-empirical formula $E_{\mathrm{coll}} = 0.2qV_0$ derived in Ref.~\citet{2005Currell}, where $q$ is the charge of the projectile.

\begin{figure}[h]
\centering
\includegraphics[width=0.85\columnwidth]{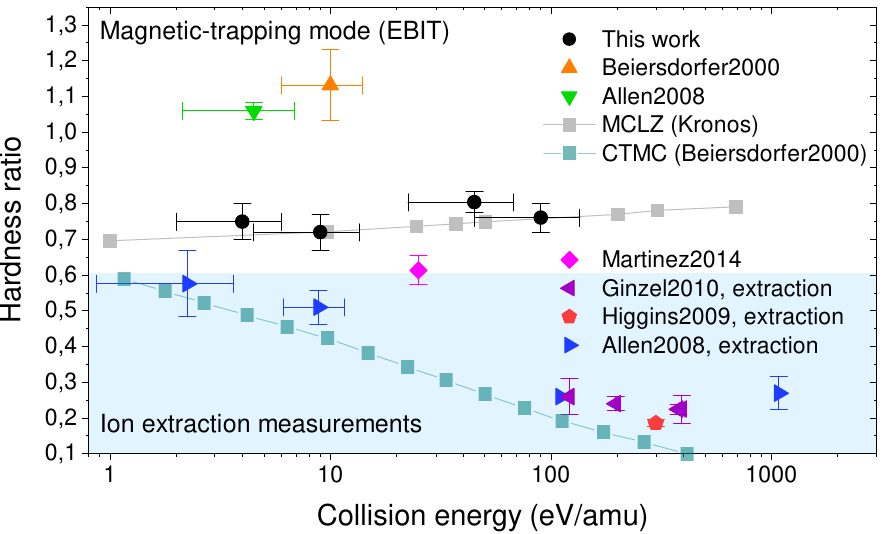}
\caption{\label{fig:ArCX_HardnessCollEnergy}Hardness ratio of $\mathrm{Ar}^{18+} + \mathrm{Ar}$ for different collision energies. All available CX measurements performed inside an EBIT are presented within the white region above $\mathcal{H} > 0.55$ and all measurements in ion-extraction experiments within the blue-shaded region. The cyan-colored curve represents the CTMC prediction~\citep{2000Beiersdorfer} and the light-grey curve the MCLZ prediction for $\mathrm{Ar}^{18+} + \mathrm{H}$ with a low-energy $\ell$-distribution.}
\end{figure}

In Fig. \ref{fig:ArCX_HardnessCollEnergy}, the experimentally obtained collision-energy-dependent hardness ratios are shown. The relative uncertainty is estimated conservatively as $50\,\%$.
Argon CX measurements in other EBITs~\citep{2000Beiersdorfer, 2008Allen, 2014Martinez} and in out-of-the-trap collision experiments~\citep{2008Allen, 2010Ginzel, 2009Higgins} are presented for comparison. It is conspicuous that the EBIT results significantly deviate from the non-EBIT ones towards higher values of the hardness ratio $\mathcal{H}$. Furthermore, $\mathcal{H}$ values from different EBITs are not in agreement with each other. %In fact, the value differs almost by a factor of two, although the experiments were performed with the same EBIT~\citep{2000Beiersdorfer, 2014Martinez} and same neutral targets. 

\begin{figure}%[htb]
\centering
\includegraphics[width=0.85\columnwidth]{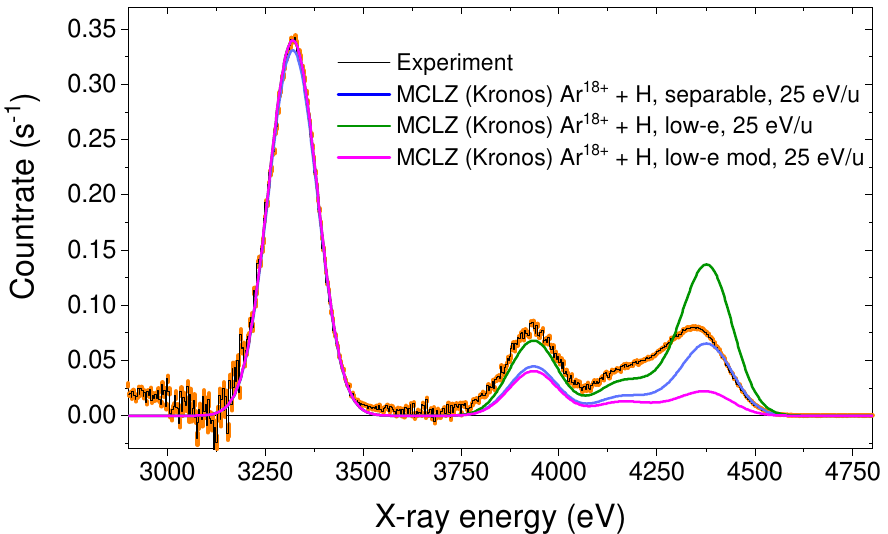}
\caption{\label{fig:ArModelCX}Experimentally obtained CX spectrum of $\mathrm{Ar}^{18+}$ interacting with a mixture of argon and residual gas, depicted as a solid curve with corresponding statistical uncertainties in orange. The colored curves represent MCLZ calculations of bare argon ions colliding with atomic hydrogen at a relative collision energy of $25\,\mathrm{eV}/\mathrm{amu}$ for different $\ell$-state distributions of the captured electrons.}
\end{figure}

\begin{figure}%[h]
\centering
\includegraphics[width=\textwidth]{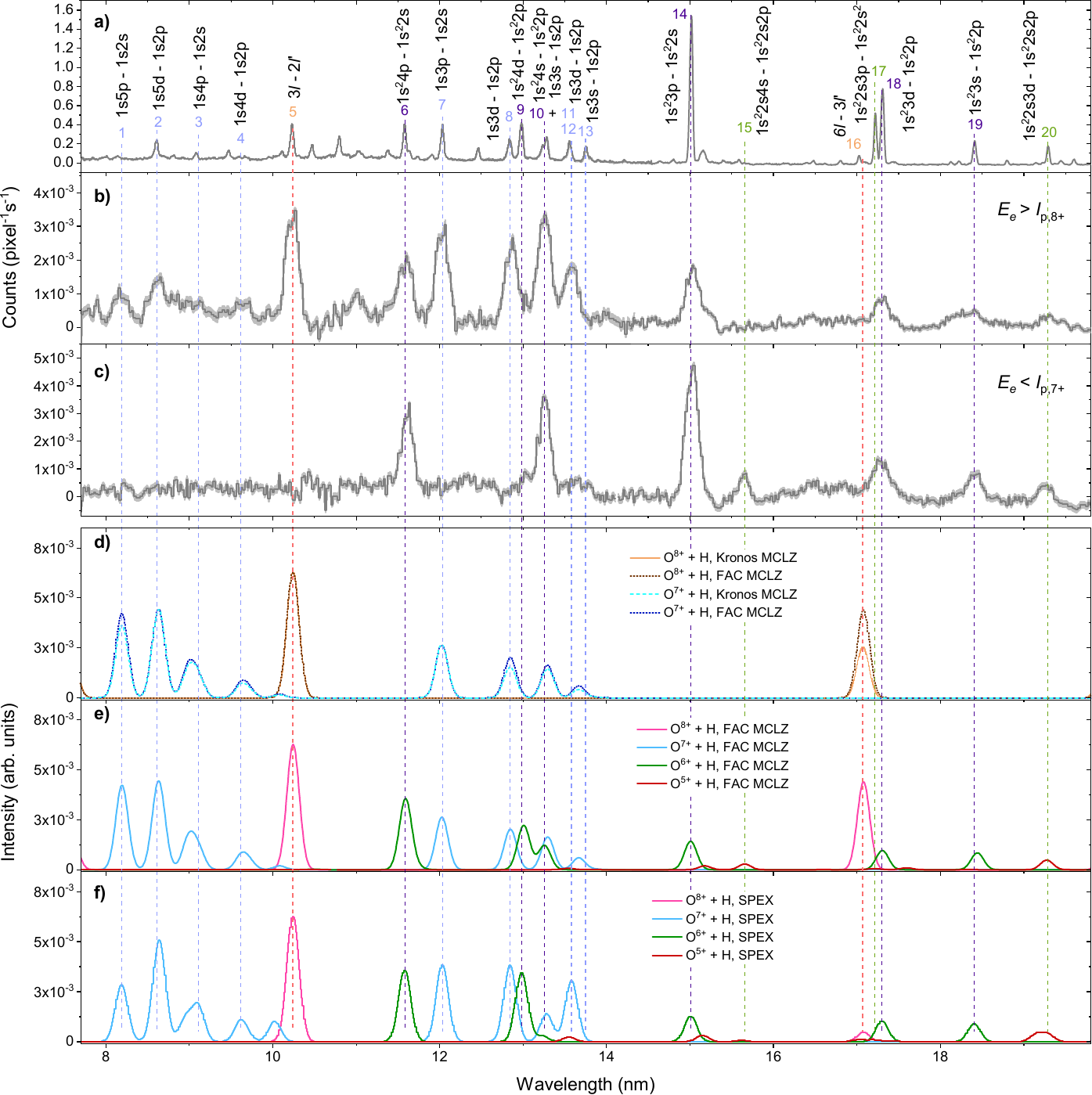}
\caption{\label{fig:O_CX_EUV}Extreme-ultraviolet spectra of highly charged oxygen ions. 
\textbf{a)}: Calibration transitions excited by electron impact with the electron beam, with identifications given in Table \ref{tab:CXLines}. 
\textbf{ b)} and \textbf{c)}: CX spectra obtained during magnetic-trapping mode for an initial electron-beam energy $E_e$ above the production threshold of fully ionized oxygen and below that of hydrogen-like oxygen, respectively. 
\textbf{d)}: Direct comparison of synthetic spectra of $\mathrm{O}^{7+} + \mathrm{H}$ and $\mathrm{O}^{8+} + \mathrm{H}$ generated with Kronos and FAC (MCLZ). 
\textbf{e)} and \textbf{f)}: Synthetic CX spectra of $\mathrm{O}^{8,7,6,5+} + \mathrm{H}$ generated with FAC (MCLZ) and with the SPEX package, respectively. 
The spectrum of each charge state is scaled to match the observed lines shown in panels b) and c). 
Note that CX cross-sections $\sigma$ from Kronos~\citep{2016Mullen} and FAC~\citep{2008Gu} codes are based on the MCLZ method, while the SPEX package~\citep{2016Gu} uses various sources of cross sections $\sigma$ for different ions.
%A collision velocity of $50\,\mathrm{km}/\mathrm{s}$ was used for the calculations.
}
\end{figure}

In our experiment, influences of the different magnetic-field strengths in the center of the trap could be excluded since no significant difference of $\mathcal{H}$ was observed by reducing the field strength from $6\,\mathrm{T}$ to $3\,\mathrm{T}$. Similar observations are reported in Ref.~\citet{2008Allen}. \revchange{Other effects remain possible. The neutral composition and density in the trap, the charge-state distribution during the breeding stage, the ion-energy distribution after beam switch-off, and the treatment of unresolved spectral blends can all modify the fitted value of $\mathcal{H}$. In addition, multi-electron capture from many-electron targets followed by autoionization may populate lower-$n$ states than predicted by a pure single-electron-capture model. A related possibility is multiple charge exchange in the trap, for example capture by $\mathrm{Ar}^{18+}$ from $\mathrm{Ar}^{+}$ ions produced in a preceding CX event. Such an ionic donor has a larger binding energy than neutral argon and would therefore shift the most probable capture shell. Furthermore, at certain atomic distances the outgoing $\mathrm{Ar}^{+}$ ion could induce Stark shifts on the Rydberg levels, which can alter the $l$-distributions and thus the final spectrum~\cite{gu2022}.
These mechanisms are plausible systematic contributions, but the present data do not allow us to isolate one of them as the unique origin of the EBIT-to-EBIT discrepancy.}

Furthermore, we also observed that the intensity ratio of Lyman-series $n\rightarrow 1$ transitions for $n>2$ to the Ly$\alpha$ transition strongly depends on the neutral target. According to the MCLZ calculations for bare-ion collisions, only the population of $n$ is affected by the choice of the neutral target. The intensity ratios are however mostly determined by the $\ell$-state distribution. Since all measurements in this work were performed with similar trapping conditions, and thus, with comparable collision energy, the intensity ratios according to MCLZ should also be similar, in contrast to our observations with different targets. Moreover, none of the commonly used $\ell$-state distributions could reproduce the data for the interaction of $\mathrm{Ar}^{18+}$ with the mixture of neutral argon and residual gas, see Fig.~\ref{fig:ArModelCX}. 
The relative $\ell$-state population is determined by the neutral target, and not by the relative collision energy. A similar conclusion was reported in~\citet{2010Leutenegger, 2013Leutenegger}, where CX of fully ionized magnesium and phosphorus with different neutral targets was investigated with a high-resolution X-ray microcalorimeter.

\subsection{Extreme-ultraviolet spectra}
We have investigated above direct transitions from excited states with $n\geq 2\rightarrow n=1$. Other interesting cascades consist of photons of lower energies. For a complete construction of CX models, such cascades have to be studied. Typical transition energies are in the extreme-ultraviolet (EUV) domain. Spectroscopy of CX in the EUV range was performed using $\mathrm{O}^{6+}$ and $\mathrm{C}^{5+}$ ions extracted from an electron cyclotron resonance ion source, interacting with $\mathrm{H}_2\mathrm{O}$ molecules inside a reaction chamber at collision energies of $0.1\,\text{--}\,7.5\,\mathrm{keV/amu}$~\citep{Bodewits2007, Bodewits2019}. Another experiment in the EUV domain was performed with $\mathrm{Xe}^{q+}\ (q=7\,\text{--}\,9)$ ions interacting with various noble gases~\citep{Tanuma2011} utilizing a similar experimental setup. Here, we perform a proof-of-principle measurement of CX in the EUV range inside an EBIT. 
We utilized a grating spectrometer covering the range from $8\,\mathrm{nm}$ to $20\,\mathrm{nm}$, where various transitions of Be-like up to H-like oxygen ions from Rydberg states into the L-shell are predicted. Oxygen ions in bare through Li-like charge states interact with a molecular beam of oxygen. Each CX spectrum covers a range of $\sim5\,\mathrm{nm}$ at once. Those presented in Fig. \ref{fig:O_CX_EUV} are composed of two separate wavelength ranges with an overlapping region between $\sim12.5-14.0\,\mathrm{nm}$. 
\begin{sidewaystable}
\caption{\label{tab:CXLines}Identified transitions of highly charged oxygen ions from Fig. \ref{fig:O_CX_EUV}. Their wavelengths (in units of nm) are extracted from the NIST database and compared to the predictions of FAC, Kronos, SPEX, and experimental values in magnetic-trapping mode. Relative differences $\delta$FAC, $\delta$Kronos, and $\delta$SPEX to the experimental values are given in units of the experimental uncertainty $\sigma$.}
\centering{
\begin{tabular}{ccrclllllllll}
\toprule
Line & Charge state & & \hspace{-2.7em}Transition & & NIST & Exp. & FAC & $\delta$FAC & Kronos & $\delta$Kronos & SPEX & $\delta$SPEX\\

1 & He-like &\conf{1}{s}{}\conf{5}{p}{}\,$^3P$ & \hspace{-2.7em}-- &\hspace{-2.7em} \conf{1}{s}{}\conf{2}{s}{}\,$^3S$ & $8.191$\footnote[1]{Not listed in the database. The values are calculated from the theoretical energy level differences provided by NIST.} & 8.182(6) & 8.194 & $1.9$ & 8.191 & $1.5$ & 8.191 & $1.5$\\
2 & He-like & \conf{1}{s}{}\conf{5}{d}{}\,$^3D$ &\hspace{-2.7em}-- &\hspace{-2.7em}\conf{1}{s}{}\conf{2}{p}{}\,$^3P$ & $8.611(9)$ & 8.627(6) & 8.634 & $1.1$ & 8.623 & $-0.67$ & 8.642 & $2.4$ \\

3 & He-like & \conf{1}{s}{}\conf{5}{d}{}\,$^1D$ &\hspace{-2.7em}-- &\hspace{-2.7em}\conf{1}{s}{}\conf{2}{p}{}\,$^1P$ & 8.93632(1) & -- & 8.986 & -- & 8.931 & -- & 8.955 & --\\
 & & \conf{1}{s}{}\conf{4}{p}{}\,$^3P$ &\hspace{-2.7em}-- &\hspace{-2.7em}\conf{1}{s}{}\conf{2}{s}{}\,$^3S$& $9.108^{\,\text{\textcolor{red}{a}}}$ & -- & $9.108$ & -- & $9.107$ & -- & $9.107$ & --\\
4 & He-like & \conf{1}{s}{}\conf{4}{d}{}\,$^3D$ &\hspace{-2.7em}-- &\hspace{-2.7em}\conf{1}{s}{}\conf{2}{p}{}\,$^3P$ & $9.614(4)$ & 9.636(8) & 9.653 & $2.1$ & 9.646 & $1.2$ & 9.627 & $-1.2$\\
5 & H-like & \conf{3}{l}{} &\hspace{-2.7em}-- &\hspace{-2.7em} \conf{2}{l'}{} & $10.240(10)$ & 10.233(3) & 10.244 & $3.7$ & 10.244 & $3.8$ & 10.244 & $3.8$ \\
6 & Li-like & \conf{1}{s}{2}\conf{4}{p}{} &\hspace{-2.7em}-- &\hspace{-2.7em}\conf{1}{s}{2}\conf{2}{s}{} & $11.583(1)$ & 11.588(6) & 11.593 & $0.8$ & -- & -- & 11.582 & $-0.95$\\
7 & He-like & \conf{1}{s}{}\conf{3}{p}{}\,$^3P$ &\hspace{-2.7em}-- &\hspace{-2.7em}\conf{1}{s}{}\conf{2}{s}{}$\,^3S$ & $12.033^{\,\text{\textcolor{red}{a}}}$ & 12.031(4) & 12.027 & $-0.96$ & 12.033 & $0.55$ & 12.033 & $0.57$\\
8 & He-like & \conf{1}{s}{}\conf{3}{d}{}\,$^3D$ &\hspace{-2.7em}-- &\hspace{-2.7em}\conf{1}{s}{}\conf{2}{p}{}\,$^3P$ & $12.845(5)$ & 12.866(5) & 12.848 & $-3.5$ & 12.845 & $-4.1$ & 12.846 & $-4.1$ \\
9 & Li-like & \conf{1}{s}{2}\conf{4}{d}{} & \hspace{-2.7em}-- &\hspace{-2.7em}\conf{1}{s}{2}\conf{2}{p}{} & $12.984(4)$ & -- & 13.007 & -- & -- & -- & 12.984 & --\\
10 & Li-like & \conf{1}{s}{2}\conf{4}{s}{} &\hspace{-2.7em}-- &\hspace{-2.7em}\conf{1}{s}{2}\conf{2}{p}{} & $13.227(5)$ & 13.258(3) & 13.257 & $-0.75$ & -- & -- & 13.228 & $-15$\\
11 & He-like & \conf{1}{s}{}\conf{3}{s}{}\,$^3S$ &\hspace{-2.7em}-- &\hspace{-2.7em}\conf{1}{s}{}\conf{2}{p}{}\,$^3P$ & $13.281(7)$ & 13.258(3) & 13.295 & $18$ & 13.283 & $13$ & 13.283 & $13$ \\
12 & He-like & \conf{1}{s}{}\conf{3}{d}{}\,$^1D$ &\hspace{-2.7em}-- &\hspace{-2.7em}\conf{1}{s}{}\conf{2}{p}{}\,$^1P$ & $13.582(1)$ & 13.572(4) & 13.670 & $25$ & 13.586 & $3.6$ & 13.582 & $2.5$\\
13 & He-like & \conf{1}{s}{}\conf{3}{s}{}\,$^1S$ &\hspace{-2.7em}-- &\hspace{-2.7em}\conf{1}{s}{}\conf{2}{p}{}\,$^1P$ & $13.751(1)$ & -- & -- & -- & -- & -- & -- & --\\
14 & Li-like & \conf{1}{s}{2}\conf{3}{p}{} &\hspace{-2.7em}-- &\hspace{-2.7em}\conf{1}{s}{2}\conf{2}{s}{} & $15.011(2)$ & 15.026(2) & 15.007 & $-9.3$ & -- & -- & 15.010 & $-8.0$\\
15 & Be-like & \conf{1}{s}{2}\conf{2}{s}{}\conf{4}{s}{} &\hspace{-2.7em}-- &\hspace{-2.7em}\conf{1}{s}{2}\conf{2}{s}{}\conf{2}{p}{} & $15.619(7)$ & 15.646(4) & 15.656 & $2.4$ & -- & -- &  $15.619\footnote[2]{Comparably low intensity}$ & $-6.8$\\
16 & H-like & \conf{6}{l}{} &\hspace{-2.7em}-- &\hspace{-2.7em}\conf{3}{l'}{} & $17.070(10)$ & -- & 17.079 & -- & 17.074 & -- & 17.081 & --\\
17 & Be-like & \conf{1}{s}{2}\conf{2}{s}{}\conf{3}{p}{} &\hspace{-2.7em}-- &\hspace{-2.7em}\conf{1}{s}{2}\conf{2}{s}{2} & $17.217(1)$ & -- & -- & -- & -- & -- & -- & -- \\
18 & Li-like & \conf{1}{s}{2}\conf{3}{d}{} & \hspace{-2.7em}-- &\hspace{-2.7em}\conf{1}{s}{2}\conf{2}{p}{} & $17.302(8)$ & 17.283(5) & 17.305 & $4.3$ & -- & -- & 17.303 & $4.0$ \\
19 & Li-like & \conf{1}{s}{2}\conf{3}{s}{} &\hspace{-2.7em}-- &\hspace{-2.7em}\conf{1}{s}{2}\conf{2}{p}{} & $18.406(5)$ & 18.409(5) & 18.443 & $6.8$ & -- & -- & 18.406 & $-0.7$\\
20 & Be-like & \conf{1}{s}{2}\conf{2}{s}{}\conf{3}{d}{} & \hspace{-2.7em}-- &\hspace{-2.7em}\conf{1}{s}{2}\conf{2}{s}{}\conf{2}{p}{} & $19.291(1)$ & 19.254(8) & 19.275 & $2.6$ & -- & -- & 19.219 & $-4.4$ \\
\bottomrule
\end{tabular}}
\end{sidewaystable}
In panel a) of Fig. \ref{fig:O_CX_EUV}, we show well-known spectral lines of various charge states of oxygen ions excited by electron impact during the beam-on period which are used for calibration. The electron beam of $170\,\mathrm{mA}$ had an energy of $1350\,\mathrm{eV}$, and an axial trapping potential of $500\,\mathrm{V}$ was used. 
From the H-like system, only two spectral lines at $10.24\,\mathrm{nm}$ and $17.07\,\mathrm{nm}$ (lines 5 and 16) are seen, namely $n=3 \to 2$ and $n=6 \to 3$, respectively. 
Panel b) shows the corresponding CX spectrum in the MTM at the same EBIT conditions. 

Note that the spectral lines in our slitless-imaging EUV spectrometer appear broader in absence of the trapping by the electron beam during the MTM. This leads to an expansion of the ion cloud by a factor of $\sim4.5$, which we also monitor using the imaging capabilities of an additional 3-m normal-incidence spectrometer installed at FLAH-EBIT. 
The initial electron-beam energy $E_{e}$ is above the production  threshold $I_{\mathrm{p},8+} \approx 871\,\mathrm{eV}$. 
Here, from the H-like system, only Balmer-$\alpha$ (line 5) ($n=3\to2$) was observed. In panel c), the electron-beam energy was tuned to $730\,\mathrm{eV}$, below the production threshold $I_{\mathrm{p},7+}$ ($\approx 739.33$ eV) for H-like oxygen ions. 
As expected, after the electron capture, we see neither hydrogenic (5 and 16) nor helium-like lines.
Furthermore, even below $I_{\mathrm{p},7+}$ a certain fraction of H-like ions is produced in two steps: first, populating He-like metastable state ($1s 2s$) by electron impact, and second, ionization by a second electron impact. This was checked by simultaneously observing X-ray K$\alpha$ photons from the metastable (line $z$ in Gabriel's notation) and their decay after the electron beam is switched off. In the EUV spectra, the He-like metastable contributions are negligible, and the observed lines on panel c) arise from Li-like and Be-like oxygen ions.
\subsection{Comparison of EUV spectra with predictions}
For the comparison between our experiments and theory, we generated synthetic spectra using various CX modeling codes, where we used H as a target and the collision velocity of $50\,\mathrm{km}/\mathrm{s}$. 
\revchange{The EUV data provide constraints that are complementary to the X-ray hardness ratios. Whereas the X-ray spectra mostly test the final direct decays into the K shell, the EUV lines probe intermediate cascade steps from high-lying Rydberg states into the L shell and are therefore sensitive to the $n\ell$ distribution after capture, to cascade branching, and to possible satellite contributions from multi-electron capture.}
Panel d) shows modeled CX spectra for initial O$^{8+}$ and O$^{7+}$ ions, generated using Kronos~\citep{2016Mullen} and FAC~\citep{2008Gu} codes.  
Both hydrogenic lines 5 and 16 are predicted by Kronos and FAC. 
However, the spectral line 16 at $17.07\,\mathrm{nm}$ was not observed in this experiment. Therefore, we cannot confirm MCLZ predictions from FAC and Kronos of electron capture into $n=6$. 
Furthermore, the solid orange curve (Kronos) and the dotted brown curve (FAC) in panel d) predict intensities of line 16 differing by almost a factor of 2.  
We normalized all models to line 2 and found out that the disagreement between FAC and Kronos is smaller for the heliumlike spectrum than for the hydrogenic one (see cyan dashed line and dotted navy blue line in panel d) for Kronos and FAC, respectively). 
\revchange{The sizable difference between Kronos-MCLZ and FAC-MCLZ, although both are based on the MCLZ approximation, is not unexpected. The MCLZ transition probabilities depend on the positions and couplings of avoided crossings, which are determined by the adopted level energies and by the set of final states included in the calculation. FAC calculates the relevant level structure and cascade matrix mostly \textit{ab initio}, while Kronos uses tabulated energies where available, mostly from NIST ASD~\cite{NIST}, quantum-defect extrapolations for high-$n$ levels, and a different cascade implementation. Small differences in the distribution among neighboring capture shells can be amplified in the EUV because several cascade branches feed the same observed lines. }
%Thus, the factor-of-two difference for the predicted O$^{8+}$ intensity in line 16 should be regarded as an implementation-dependent uncertainty of present MCLZ-based modeling rather than as a contradiction between two otherwise identical calculations.}
Besides hydrogenic and helium-like CX transitions, the Kronos model provides no data for the lower charge states of oxygen ions.
Thus, we used FAC to calculate synthetic spectra for lithiumlike and beryliumlike ions after the electron capture. 
Panel e) shows them as solid red and green curves. 
Lines 6, 10, 14, 15, 18, 19, and 20, which appear in panel b) and c), are identified. Except lines 15 and 20, all lines are due to CX between O$^{6+}$ and hydrogen. 

In panel f), we also modeled the EUV spectra using the SPEX package. 
Here, the theoretical cross sections are taken mostly from~\citet{1993Janev} for $\mathrm{O}^{8+}$, and \citet{2012Wu} for $\mathrm{O}^{6+}$ ions. 
Note that the calculations by Nolte and Wu are done with the QMMOCC method, while the other two are recommended values based on data compilations. 
We then normalized the model to our data for the strongest lines of each ion. 
The line width is adjusted to fit the line profiles of the main transitions. 
Although we found some agreement between FAC- and SPEX-generated synthetic spectra, both models underestimate the intensity of line 10 at $13.2\,\mathrm{nm}$ and overestimate the line at $13.0\,\mathrm{nm}$ of the lithium-like charge state after electron capture. Furthermore, we do not see an increase of the intensity at the position of line 9 in MTM (panel c) of Fig. \ref{fig:O_CX_EUV}), where FAC and SPEX predict an intense lithium-like transition.
We note that the AtomDB-ACX model does not predict any transition in the EUV domain. 

In Tab. \ref{tab:CXLines}, the transition energies predicted by FAC, Kronos, and SPEX as well as the experimental results in magnetic-trapping mode are summarized. 
The differences in transition wavelengths $\delta$FAC, $\delta$Kronos, and $\delta$SPEX between the experiment and the respective model are given in units of the experimental uncertainty, where negative values denote a shift towards lower wavelengths than observed in the experiment.
As can be seen in Tab. \ref{tab:CXLines}, only a few lines are predicted by each of the codes. Therefore, we cannot estimate the respective uncertainties of the modeled spectra, or compare their reliability. 

In general, the modeled CX spectra rely on the $n\ell S$-resolved CX cross sections calculated by different theories and the specific radiative cascades model used in the codes. 
The accuracy of CX cross sections depends on the model. While, e. g., the MCLZ method is the convenient for calculations, it uses more approximations than methods such as atomic orbital close coupling (AOCC) and quantum mechanical molecular orbital close coupling (QMOCC). Moreover, the treatment of radiative cascades models also shows differences. Inherent uncertainties exist for the radiative rates and autoionization rates compiled in the models. In order to reduce the total cascade matrix and accelerate the modeling, some codes such as Kronos use only dipole radiative decays and ignore higher-order multipole channels. 
The SPEX code uses reliable cross sections from AOCC and QMOCC theories when available in the literature, but scaling relations to estimate missing cross-section data. 
While FAC also uses the MCLZ method, as Kronos does, to calculate the CX spectrum, it has the advantage  that it calculates all atomic data \textit{ab initio}. Thus, FAC provides better control of the sets of configurations, maximum higher-$n$ quantum numbers (while Kronos relies on quantum defect method to calculate high-$n$ energy levels), and includes higher-order multipoles and two-photon decay rates in the calculations. 
As the resulting MCLZ cross sections strongly depend on the energy level separations, MCLZ cross sections and cascade models can be improved by expanding the configuration basis set in FAC to obtain more reliable level energies. 

A general problem is that these standard codes only include single-electron capture (SEC). \revchange{Multi-electron capture (MEC), in which two or more electrons are transferred, can be strong in collisions with neutrals with more than one electron, as shown by \citet{2005Ali}. MEC followed by autoionization can leave the same final charge state as SEC but with a different population of the radiating levels, which is one possible route to the enhanced lower-$n$ emission seen in the EUV spectra.} 
In Tab. \ref{tab:Methods}, the different codes and methods used to generate synthetic CX EUV spectra for this work are summarized. Note that Kronos also provides a QMMOCC calculation for the cross section of the $\mathrm{O}^{7+}+\mathrm{H}$ collision, but at higher collision energies ($>200\,\mathrm{eV}/\mathrm{amu}$) than the one used in this work, and thus we do not compare it with the experiment. 
\begin{table*}
\begin{center}
\caption{\label{tab:Methods}Summary of the different codes and methods applied for the generation of synthetic CX spectra in this work. The last two columns show the sources energy levels and CX cross sections we used. RV refers to recommended values based on data compilations.}
\centering{
\begin{tabular}{ccccc}
%\begin{tabular}{cr@{\,--\,}lc}
\toprule
Ion&Method&Type&Energy levels&Ref. for $\sigma_{\mathrm{CX}}$\\
%Label & \multicolumn{2}{c}{Transition} & \mathrm{NIST wavelength}$\,(\mathrm{nm})$\\
%Label & Transition & & NIST wavelength$\,(\mathrm{nm})$\\
\midrule

$\mathrm{O}^{8+}$&MCLZ&FAC&ab initio&\cite{2008Gu}\\
$\mathrm{O}^{8+}$&MCLZ&Kronos&NIST&\cite{2016Mullen}\\
$\mathrm{O}^{8+}$&RV&SPEX& &\cite{1993Janev}\\
\midrule
$\mathrm{O}^{7+}$&MCLZ&FAC&ab initio&\cite{2008Gu}\\
$\mathrm{O}^{7+}$&MCLZ&Kronos&NIST&\cite{2016Mullen}\\
$\mathrm{O}^{7+}$&QMMOCC&SPEX& & --\\
\midrule
$\mathrm{O}^{6+}$&MCLZ&FAC&ab initio&\cite{2008Gu}\\
$\mathrm{O}^{6+}$&QMMOCC&SPEX& &\cite{2012Wu}\\
\midrule
$\mathrm{O}^{5+}$&MCLZ&FAC&ab initio&\cite{2008Gu}\\
$\mathrm{O}^{5+}$&RV&SPEX& &-- \\
\bottomrule
\end{tabular}}
\end{center}
\end{table*}

\section{Conclusions and outlook}

We presented charge-exchange measurements of fully ionized argon with different neutral donors performed at similar conditions inside an electron beam ion trap. No obvious scaling of the hardness ratio with the ionization potential of the neutral target was found. In contrast, the SEC MCLZ model predicts a steady increase with higher ionization energies. \revchange{One plausible, but not uniquely established, explanation for this disagreement is two-electron capture followed by autoionization of one electron. Such a pathway would lower the most probable $n_{\mathrm{CX}}$ of the captured electron while leading to the same final charge state as single-electron capture.} Additionally, extensive studies of the hardness ratio of the $\mathrm{Ar}^{18+} + \mathrm{Ar}$ interaction were performed at different axial trapping potentials, and thus, at different relative collision energies. Within the statistical uncertainties, no significant energy dependence was observed. MCLZ predicts an increase of the hardness ratio with collision energy, while CTMC predicts a decrease. 

Measurements conducted with EBITs differ significantly from each other, although experimental conditions were comparable. 
Furthermore, the hardness ratios determined in EBIT measurements differ significantly towards higher values from measurements of ions interacting with neutral argon in a gas cell. Possible systematic effects are discussed in Ref. \cite{2008Allen}, but so far, no conclusive explanation could be provided. The CX spectrum of $\mathrm{Ar}^{18+}$ interacting with neutral argon and residual gas could not be represented by MCLZ calculations with different applied $\ell$-state distributions. \revchange{We therefore regard MEC, multiple CX involving ionic donors, unresolved blends, detector response, and trap-specific ion-energy distributions as possible contributions rather than as individually proven explanations. In particular, CX between $\mathrm{Ar}^{18+}$ and $\mathrm{Ar}^{+}$ ions produced in preceding CX collisions cannot be excluded and would tend to lower $n_{\mathrm{CX}}$ because the donor electron is more strongly bound than in neutral argon.}

The discrepancy between MCLZ calculations and CX experiments is even more prominent in the EUV domain, where transitions from highly excited states into the L-shell of hydrogen-like oxygen after electron transfer have been studied. Predicted capture into $n=6$ could not be confirmed by our experiment. For helium-like oxygen ions after electron capture, the predicted capture into $n=5$, exclusively, was also not observed. Transitions from $n=3$ to $n=2$ are much underestimated by the model. \revchange{These EUV observations provide more information than the X-ray hardness ratios because they directly show which cascade branches into the L shell are over- or underpopulated. They therefore constrain not only the total hard-to-soft X-ray balance, but also the redistribution among neighboring $n$ and $\ell$ states during the cascade.} The calculation based on the SPEX-CX package with theoretical QMMOCC cross sections agrees better with our experiment. \revchange{A possible, still not definitive, explanation for a remaining mismatch is multi-electron transfer followed by autoionization, which would enhance lower-$n$ populations without changing the final charge state.}

In conclusion, our present X-ray and EUV spectral data on CX by argon and oxygen ions \revchange{can be} readily used for comparison with astrophysical observations. However, a clear theoretical picture \revchange{emerges neither from this work nor from other experimental studies}. The disagreements seen in experiments running at comparable conditions \revchange{point to dependencies of the CX process on experimental conditions that are not yet well understood}. In order to clarify the situation, more systematic laboratory studies are required. Besides EBIT experiments, COLTRIMS and merged-beam setups were used to understand the kinematics of CX in detail and extract state-resolved differential cross sections (e.~g., in ~\citep{2002Fischer,Knoop_2008,Kelkar_2012,Xue2014}) for single-electron capture. With such techniques, combined with spectroscopic instruments, it could be possible to understand the complete kinematics of multi-electron capture~\citep{2005Ali, 2010Ali, 2016Ali} since it is possible to track the photons and ions before and after the electron transfer in coincidence. On the other hand, to avoid possible contributions from multi-electron capture, atomic hydrogen is required as a target in future CX experiments, where only single-electron capture can occur. Furthermore, its abundance in the Universe requires such laboratory data for the analysis of upcoming high-resolution astrophysical observations. The interpretation of astrophysical spectra of space missions such as \textit{XRISM}~\citep{Tashiro2022XRISM,Tashiro2025XRISM,Tashiro2024XRISMstatus} or the future \textit{ATHENA}~\citep{2013Barret,Cruise2025NewAthena} urgently calls for improvements of CX models, more reliable $n\ell$-resolved cross-sections, and benchmarking through laboratory experiments.

\bmhead{Data availability} Those results in this work that were derived using simulations can be reproduced as detailed in the text. Original data sets generated during the current study are available from the corresponding author on reasonable request.

\bmhead{Acknowledgements}
This work was supported by the Max-Planck-Gesellschaft (MPG). C.S. acknowledges support from the MPG, the Deutsche Forschungsgemeinschaft (DFG) under Project No. 266229290, and the NASA--JHU Cooperative Agreement. SRON is supported financially by NWO, the Netherlands Organization for Scientific Research.The research leading to these results has received funding from the European Union's Horizon 2020 Programme under the AHEAD2020 project (grant agreement n. 871158 and n. 654215)

\bibliography{biblio2_R2_marked}
\end{document}